\documentclass[prl, twocolumn, superscriptaddress]{revtex4}

\usepackage{amsmath} 
\usepackage{amsfonts}
\usepackage{graphicx}  
\usepackage{verbatim}  
\usepackage{hyperref}  

\usepackage{color}
\newcommand{\V}[1]{\ensuremath{\boldsymbol{ #1}}}

\newcommand{\Vh}[1]{\ensuremath{\hat{\boldsymbol{ #1}}}}

\newcommand{\eqnref}[1]{eqn.~\ref{#1}}
\newcommand{\figref}[1]{Fig.~\ref{#1}}

\begin{document}

\title{Helicity conservation by flow across scales in reconnecting vortex links and knots}

\author{Martin W. Scheeler}
\thanks{These two authors contributed equally.}
\affiliation{James Franck Institute, Department of Physics, The University of Chicago, Chicago, Illinois 60637, USA.}
\author{Dustin Kleckner$^*$}
\email{dkleckner@uchicago.edu}
\affiliation{James Franck Institute, Department of Physics, The University of Chicago, Chicago, Illinois 60637, USA.}
\author{Davide Proment}
\affiliation{School of Mathematics, The University of East Anglia, NR4 7TJ Norwich, Norfolk, UK.}
\author{Gordon L. Kindlmann}
\affiliation{Computation Institute, Department of Computer Science, The University of Chicago, Chicago, Illinois 60637, USA.}
\author{William T.M. Irvine}
\email{wtmirvine@uchicago.edu}
\affiliation{James Franck Institute, Department of Physics, The University of Chicago, Chicago, Illinois 60637, USA.}

\begin{abstract}
The conjecture that helicity (or knottedness) is a fundamental conserved  quantity has a rich history in fluid mechanics, but the nature of this conservation in the presence of dissipation has proven difficult to resolve.
Making use of recent advances, we create vortex knots and links in viscous fluids and simulated superfluids and track their geometry through topology changing reconnections.
We find that the reassociation of vortex lines through a  reconnection  enables the  transfer of helicity from  links and knots to helical coils.
This process is remarkably efficient, owing to the  anti-parallel orientation spontaneously adopted by the reconnecting vortices.
Using a new method for quantifying the spatial helicity spectrum, we find that the reconnection process can be viewed as transferring helicity between scales, rather than dissipating it.
We also infer the presence of geometric deformations which convert helical coils into even smaller scale twist, where it may ultimately be dissipated.
Our results suggest that helicity conservation plays an important role in fluids and related fields, even in the presence of dissipation.
\end{abstract}

\maketitle

In addition to energy, momentum and angular momentum, ideal (Euler) fluids have an additional conserved quantity -- helicity (\eqnref{eqn:H}) -- which measures the linking and knotting of the vortex lines composing a flow~\cite{Moffatt1969a}.
For an ideal fluid, the conservation of helicity is a direct consequence of the Helmholtz laws of vortex motion, which both forbid vortex lines from ever crossing and preserve the flux of vorticity, making it impossible for linked or knotted vortices to ever untie~\cite{Moffatt1969a, Thomson1867a}.
Since conservation laws are of fundamental importance  in  understanding  flows, the question of  whether  this topological conservation law  extends to real, dissipative systems is of clear and considerable interest.
The general importance of this question is further underscored by the recent and growing impact knots and links are having across a range of fields, including plasmas~\cite{Moffatt2014,Ricca1996}, liquid crystals~\cite{Tkalec2011, Martinez2014},  optical~\cite{Dennis2011}, electromagnetic~\cite{Kedia2013} and biological structures~\cite{Han2010,Chichak2004, Sumners1995},  cosmic strings~\cite{Vachaspati1994,Bekenstein1992} and beyond~\cite{Faddeev1997}. Determining if and how helicity is conserved in the presence of dissipation is therefore paramount in understanding the fundamental dynamics of real fluids and the connections between tangled fields across systems.

The robustness of helicity conservation in real fluids is unclear because dissipation allows the topology of field lines to change.
For example, in viscous flows vorticity will diffuse, allowing nearby vortex tubes to `reconnect' (\figref{fig:reconnection}{\bf A-C}), creating or destroying the topological linking of vortices.
This behavior is not unique to classical fluids: analogous reconnection events have also been experimentally observed in superfluids~\cite{Bewley2008} and coronal loops of plasma on the surface of the sun~\cite{Cirtain2013}.
In general, these observed reconnection events exhibit divergent, non-linear dynamics which makes it difficult to resolve helicity dynamics theoretically~\cite{Ricca1996,KIMURA2014,Kida1988a}.
On the other hand, experimental tests of helicity conservation have been hindered by the lack of techniques to create vortices with topological structure.
Thanks to a recent advance~\cite{Kleckner2013}, this is finally possible.

By performing experiments on linked and knotted  vortices  in water, as well as  numerical simulations of Bose-Einstein condensates (a compressible superfluid~\cite{Pitaevskii2003}) and  Biot-Savart vortex evolution, we investigate the conservation of helicity, in so far as it can be inferred from  the centerlines of reconnecting vortex tubes.
We  describe a new method for quantifying the storage of helicity on different spatial scales of a thin-core vortex: a `helistogram'.
Using this analysis technique, we find  a rich structure in the flow of helicity, in which geometric deformations and vortex reconnections transport helicity between scales.
Remarkably, we find that helicity can be conserved even when vortex topology changes dramatically, and identify a system-independent geometric mechanism for efficiently converting helicity from links and knots into helical coils.

\section{Topology and Helicity}

\begin{figure}
\includegraphics{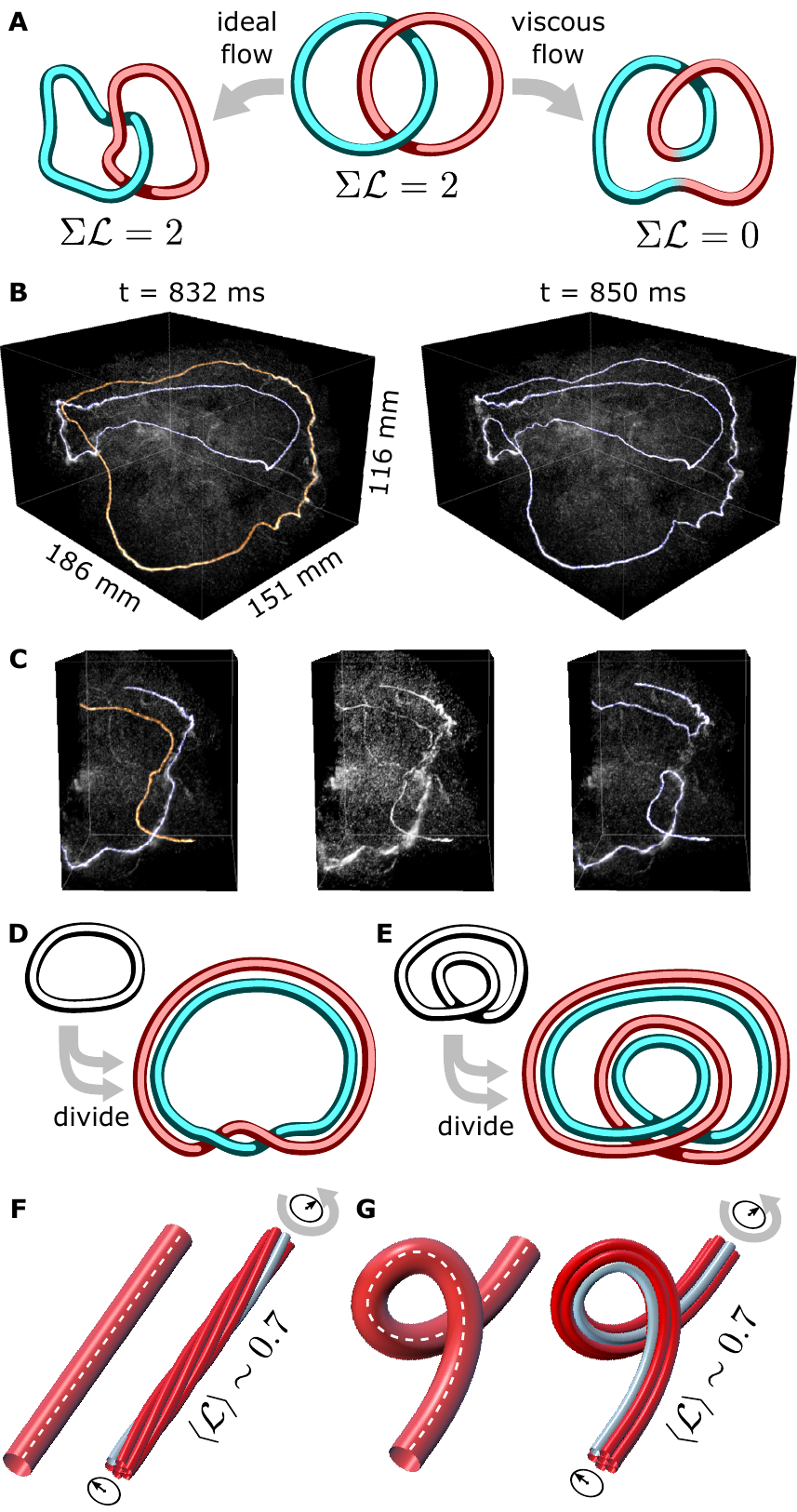}
\caption{
\label{fig:reconnection}
({\bf A}) A sketch of the evolution of vortex tube topology in ideal (Euler) and viscous (Navier-Stokes) flow.  Dissipative flows allow for reconnections of vortex tubes, and so \emph{tube} topology is not conserved.
({\bf B}) Two frames of a 3D reconstruction of a vortex reconnection in experiment, which turns an initially linked pair of rings into a single twisted ring.
({\bf C}) A close-up view of the reconnection in {\bf B}.
({\bf D}) If a tube is subdivided into multiple tubes, linking between the two may be created by introducing a twist into the pair.
({\bf E}) Similarly, if a coiled tube is subdivided, linking can result even without adding twist.  This can be seen either by calculating the linking number for the pair, or imagining trying to separate the two.
({\bf F}) In a continuum fluid, the vortex tube may be regarded as a bundle of vortex filaments, which may be twisted.
In this case a twist of $\Delta \theta \sim 0.7 \times 2\pi$ results in a total helicity of $\mathcal{H} \sim 0.7 \Gamma^2$.
({\bf G}) If the vortex tube is coiled, linking will also be introduced, as in {\bf E}.  Conceptually, this coiling can be regarded as producing a net rotation of the vortex bundle even when it is everywhere locally untwisted.
}
\end{figure}

Topology in a fluid is stored in the linking of vortex lines.
The simplest example of linked vortex lines is a joined pair of  rings (\figref{fig:reconnection}{\bf A}); however, the same topology can be obtained with very different geometries, for example by twisting or coiling a pair of rings (\figref{fig:reconnection}{\bf DE}).

Vortex loops in fluids (e.g. \figref{fig:reconnection}{\bf BC}) consist of a core region of concentrated vorticity, $\V \omega$, that rotates around the vortex centerline, surrounded by irrotational fluid motion.
In the language of vortex lines, vortices should therefore be regarded as `bundles' of vortex `filaments'  (e.g. \figref{fig:reconnection}{\bf FG}), more akin to stranded rope than an infinitesimal line.
In this case, topology can be stored either by linking and knotting of bundles, or by linking of nearby filaments within a single bundle.

The hydrodynamic helicity quantifies the degree of vortex linking present in a flow; in terms of the flow field, $\V u(\V r)$ (where $\V r$ is the spatial coordinate), it is given by:
\begin{equation}
\mathcal{H} = \int \V u \cdot \V \omega\ d^3 r,  \label{eqn:H}
\end{equation}
where the vorticity is $\V \omega = \V \nabla \times \V u$.
This quantity is exactly conserved for ideal fluids~\cite{Moffatt1969a, Moreau1961}.
The connection between helicity and the linking between vortex tubes was first noted by Moffatt~\cite{Moffatt1969a}, who showed that for a flow consisting of thin, closed vortex lines $\mathcal C_n$, the helicity is equivalently given by:
\begin{equation}
\mathcal{H} = \sum_{i, j} \Gamma_i \Gamma_j \frac{1}{4\pi} \oint \limits_{\mathcal C_i} \oint  \limits_{\mathcal C_j}\frac{\V x_i - \V x_j }{|\V x_i - \V x_j|^3} \cdot (d\V \ell_i\times d\V \ell_j), \label{eqn:GL}
\end{equation}
where $\Gamma_i$ and $\V x_i$ correspond to the circulation (vorticity flux) and  path of vortex tube $\mathcal C_i$.
The resulting double path integral was recognized as the Gauss linking integral, which measures the linking between the paths $C_i$ and $C_j$ or in the case $i=j$ the writhe (coiling and knotting) of a single path.

For finite thickness vortex tubes, one may sub-divide the bundle into $N$ infinitesimal filaments each with strength $\Gamma/N$, and compute \eqnref{eqn:GL} in the limit $N\rightarrow \infty$~\cite{Berger2006, Arnold1986}.
The result is conveniently expressed as the sum of three terms, each geometrically distinct contributions to the same measure of topology:
\begin{equation}
\mathcal{H} =\sum_{i \neq j}  \Gamma_i \Gamma_j\ \mathcal L_{ij}+ \sum_{i} \Gamma_i^2\  \left(Wr_i + Tw_i\right), \label{eqn:LkWrTw}
\end{equation}
where $\mathcal L_{ij}$ is the linking number between bundles $i$ and $j$,  $Wr_i$ is the writhe of the bundle centerline and $Tw_i$ is  the  total twist of each bundle.
Both the linking number and writhe are given by the Gauss linking integral for the bundle centerline.
The writhe of a curve quantifies its total helix-like coiling and knotting.
The twist is given by \hbox{$Tw = \frac{1}{2\pi}\oint (\Vh{n} \times \partial_s \Vh{n}) \cdot d\V \ell$}, where $\Vh n$ is a normal vector on each path which describes the bundle orientation.

The first sum of \eqnref{eqn:LkWrTw}, for $i\neq j$, measures the linking between bundles, while the second sum, over $i$, measures the linking between filaments within each bundle.
The topological contribution of twist can be visualized by sub-dividing a ring (with flux $\Gamma$) into a pair of  filaments (with flux $\Gamma/2$), that twist around each other as shown in \figref{fig:reconnection}{\bf D}. Similarly, the topological contribution of writhe can be seen in  \figref{fig:reconnection}{\bf E}; in each case the resulting helicity is $\mathcal H = 1 \Gamma^2$.
Each of these examples produces a helicity equal to an integer multiple of $\Gamma^2$, however, for a bundle the helicity is a flux-weighted average linking which need not be an integer multiple of $\Gamma^2$ (e.g., if a filament does not close in a single trip around  the bundle)~\cite{Arnold1986}.
The bundle sections shown in \figref{fig:reconnection}{\bf FG} each have a helicity of $\mathcal H \sim 0.7\Gamma^2$, resulting from  twist or  coil, respectively.
The topological equivalence of these two geometrically distinct bundles can be seen by taking the coiled bundle and pulling on the ends; straightening out the coil results in a compensating twist, which conserves the helicity.

While the linking number between bundles and the writhe of each bundle can be calculated from the centerline alone, measuring the twist requires additional information about the fine structure of the vortex core, which is challenging to resolve experimentally.
For the remainder of the manuscript we consider only the `centerline helicity', given by:
\begin{equation}
\mathcal H_c/\Gamma^2 = \sum_{i \neq j} \mathcal L_{ij} + \sum_i Wr_i. \label{eqn:Hc}
\end{equation}
This geometric quantity is equivalent to the total helicity if we assume that all vortex tubes have the same circulation, $\Gamma_i \rightarrow \Gamma$, and are locally untwisted, $Tw_i=0$.
We note that twist is naturally dissipated by viscosity, and for a twisted straight line vortex filament, this occurs at a rate $\frac{\partial_t Tw}{Tw} = -\frac{8 \pi \nu}{A_{\rm eff}}$, where $\nu$ is the kinematic viscosity and $A_{\rm eff}$ is the bundle cross sectional area (see supplementary materials).
For typical experimental parameters this can be estimated to be faster than the overall dynamics of the centerline motion.

\section{Methods}

To explore the behavior of thin-core vortices in experiment, we create shaped vortex loops in water, akin to the familiar smoke ring, but imaged with buoyant micro-bubbles instead of smoke.
Our vortex loops are generated by impulsively accelerating specially shaped, 3D printed hydrofoils.
Upon acceleration, a `starting vortex'  whose shape traces the trailing edge of the hydrofoil is shed and subsequently evolves under its own influence; using this technique it is possible to generate arbitrary geometry and topology, including links and knots~\cite{Kleckner2013}.
In order to study the effects of topology, we focus on the behavior of the most elemental linked and knotted vortices: Hopf links and trefoil knots, both of which can be created with high fidelity.
Our vortices have a typical width of 150 mm and circulation of $\Gamma= 20,000$ mm$^2/$s, and we use water as the experimental fluid.
The Reynolds number is of order $Re \sim 2\times 10^4$.

\begin{figure*}
\includegraphics{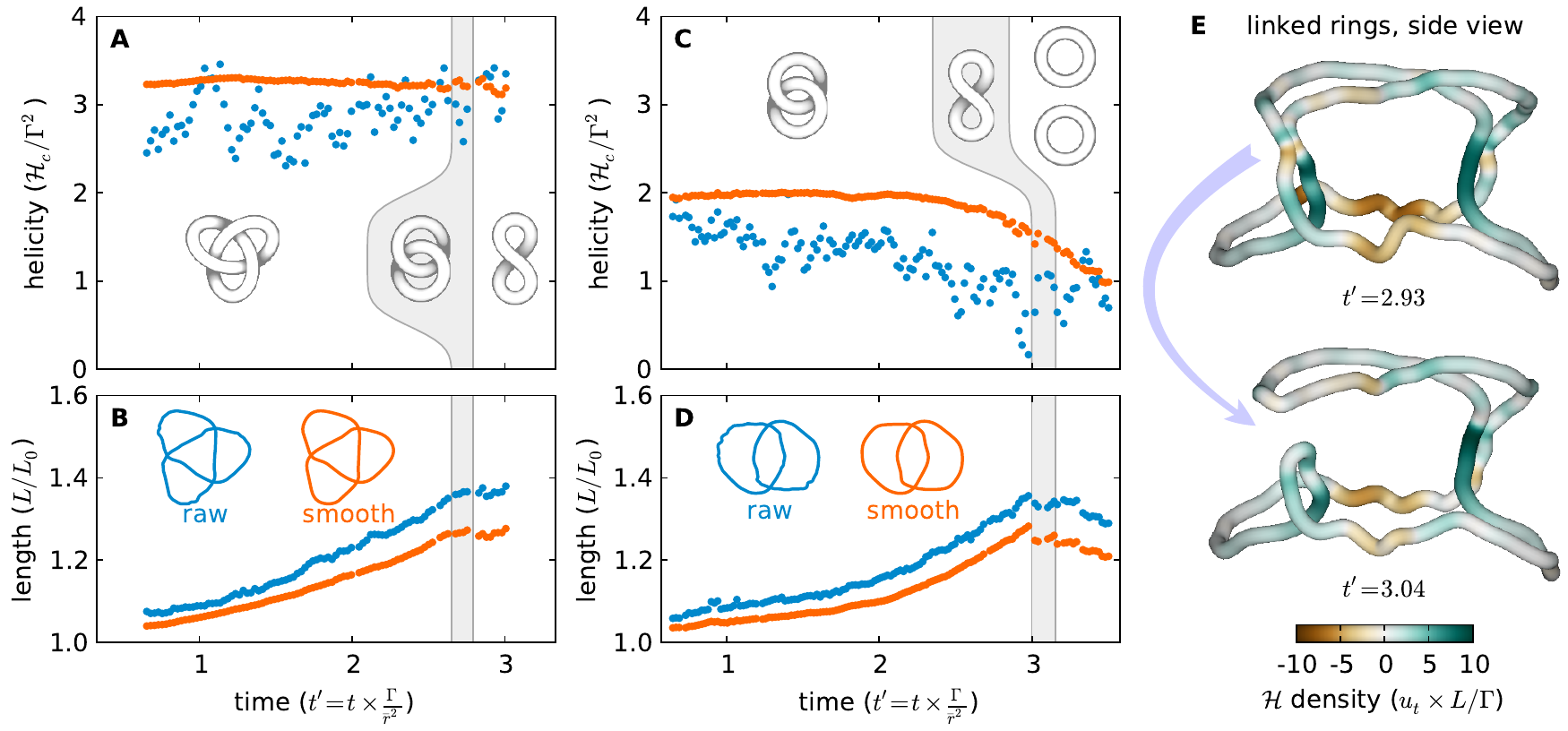}
\caption{
\label{fig:Hdata}
({\bf A, B}) The computed centerline helicity ($\mathcal H_c$) and length for of an experimental trefoil knot vortex through the first two reconnection events (out of three total), showing the efficient conversion of helicity between linking and writhing as the vortex changes topology.  The teal data indicates the raw experimental traces, while the orange data has been smoothed with a windowed sinc function whose spatial cutoff is $\lambda = 50$ mm (the total vortex length is $\sim$1 m).  The gray inset diagrams indicate the topologies at different stages of the vortex evolution.
(\hbox{\bf C, D}) The centerline helicity and length for a linked pair of vortex rings in experiment, through two reconnections.
({\bf E})  Two traces of a pair of initially linked vortices in experiment, just before and after a reconnection event.  The traces are colored according to the computed local helicity density, $h = \Gamma u_t$, calculated using the Biot-Savart law.
}
\end{figure*}

These vortices are tracked using $\sim$100 $\mu$m micro bubbles, generated by hydrolysis, which are trapped in the core of the rapidly spinning vortices (e.g., \figref{fig:reconnection}{\bf BC}).
These bubbles are in turn imaged with high speed laser scanning tomography of a $230\times230\times230$ mm volume at a resolution of $384^3$ and a rate of 170 s$^{-1}$.
Using this data, we trace the vortex cores by first identifying line-like features in the volumetric data and then connecting them to create closed 3D paths~\cite{Eberly1996, GLK:Kindlmann2009, Sethian1996}, approximated as polygons with $\sim$3,000 points.
Some disruption of the imaging and tracking results from vortex reconnections, but careful adjustment of the experimental parameters allows vortices to be tracked immediately before and after reconnections.
From these paths, physical quantities such as energy, momentum and helicity can be directly calculated as path integrals, and the geometric nature of this description allows direct comparison to other fluid systems, including simulations of superfluids and idealized thin core vortex models, both of which will be described later.
We rescale all the vortex lengths in terms of the initial length, $L_0$, and rescale the time in terms of the inital r.m.s.~vortex radius, $\bar r = \sqrt{\left<|\V x|^2\right> - |\!\left<\V x\right>\!|^2}$, and circulation, $\Gamma$: $t' = t \times \Gamma/\bar r^2$ (both $\bar r$ and $L_0$ are calculated from the \emph{designed} vortex geometry, determined by the hydrofoil shape).
Technical details for all systems are described in the supplementary methods section, and follow established methods~\cite{Kleckner2013,Proment2012, Berloff2004, Salman2013}.
Note that for our experimental vortices, we can estimate the rate of twist dissipation as $\partial_t Tw/Tw \sim 5\textrm{ s}^{-1}$, while the overall vortex motion has a timescale of order 1 s (see supplementary materials).

\section{Experimental Results}

As was found in previous studies \cite{Kida1988a,Kleckner2013,Proment2012}, our initially linked and knotted vortices disentangle themselves through local reconnections into topological trivial vortex rings.
This change in tube topology might be expected to result in a corresponding change of the helicity, since it is a global measure of the vortex topology.
For example, a reconnection event which changes a pair of linked rings into a single coiled ring (e.g. \figref{fig:reconnection}{\bf A}) should result in a sudden, discontinuous jump of the helicity by $|\Delta \mathcal{H}_c| \sim 1\ \Gamma^2$.
Recently, more detailed analytical results have also indicated that helicity may be dissipated in a reconnection event~\cite{KIMURA2014}.
Remarkably, our experimental measurements of the total centerline helicity, $\mathcal H_c$, show that it is nearly unaffected by reconnections (\figref{fig:Hdata}{\bf AC}).
As numerical computation of writhe is sensitive to small scale noise in the extracted path, applying a small amount of local smoothing to the raw path data dramatically improves the measurement.
We do this by convolving the raw vortex centerline traces, $\V x_i(s)$ (where $s$ is the path-length coordinate), with a windowed sinc function with a spatial cutoff of $\lambda = 50$ mm, which is about 5
Remarkably, we find that a vortex initially shaped into a trefoil vortex knot (\figref{fig:Hdata}{\bf AB}) is observed to have nearly constant centerline helicity, $\mathcal H_c/\Gamma^2 = 3.25 \pm 0.04$, even though it is undergoing dramatic changes in geometry and topology.
Similarly, the initially linked pair of rings (\figref{fig:Hdata}{\bf CD}) also shows no jump in the helicity through reconnection events, even though on longer timescales the centerline helicity is seen to change from $\mathcal H_c \sim 2 \Gamma^2$ to $\sim 1 \Gamma^2$, apparently via geometric deformations.
Taken together, we conclude that any jump in helicity is less than $\Delta \mathcal{H}_c \lesssim 0.05 \Gamma^2$ per reconnection for our vortices.
\begin{figure}
\includegraphics{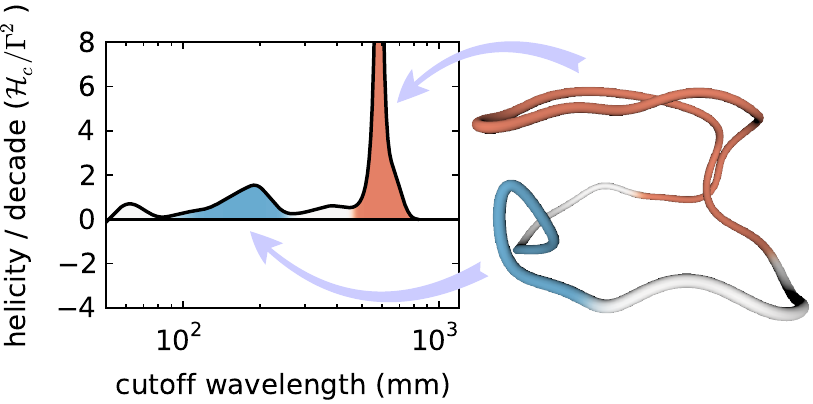}
\caption{
\label{fig:helistoexample}
The coiling component of a helistogram for an experimental pair of linked rings just after the first reconnection ($t' = 3.25$), with a colored image of the experimental data trace used to compute the helistogram.  The peaks in the helistogram correspond to coils at two different length scales, which are color coded.  In each case, the length of the segment colored is equal to the cutoff wavelength for that peak. Each coil contains approximately one unit of helicity.
}
\end{figure}
\begin{figure*}
\includegraphics{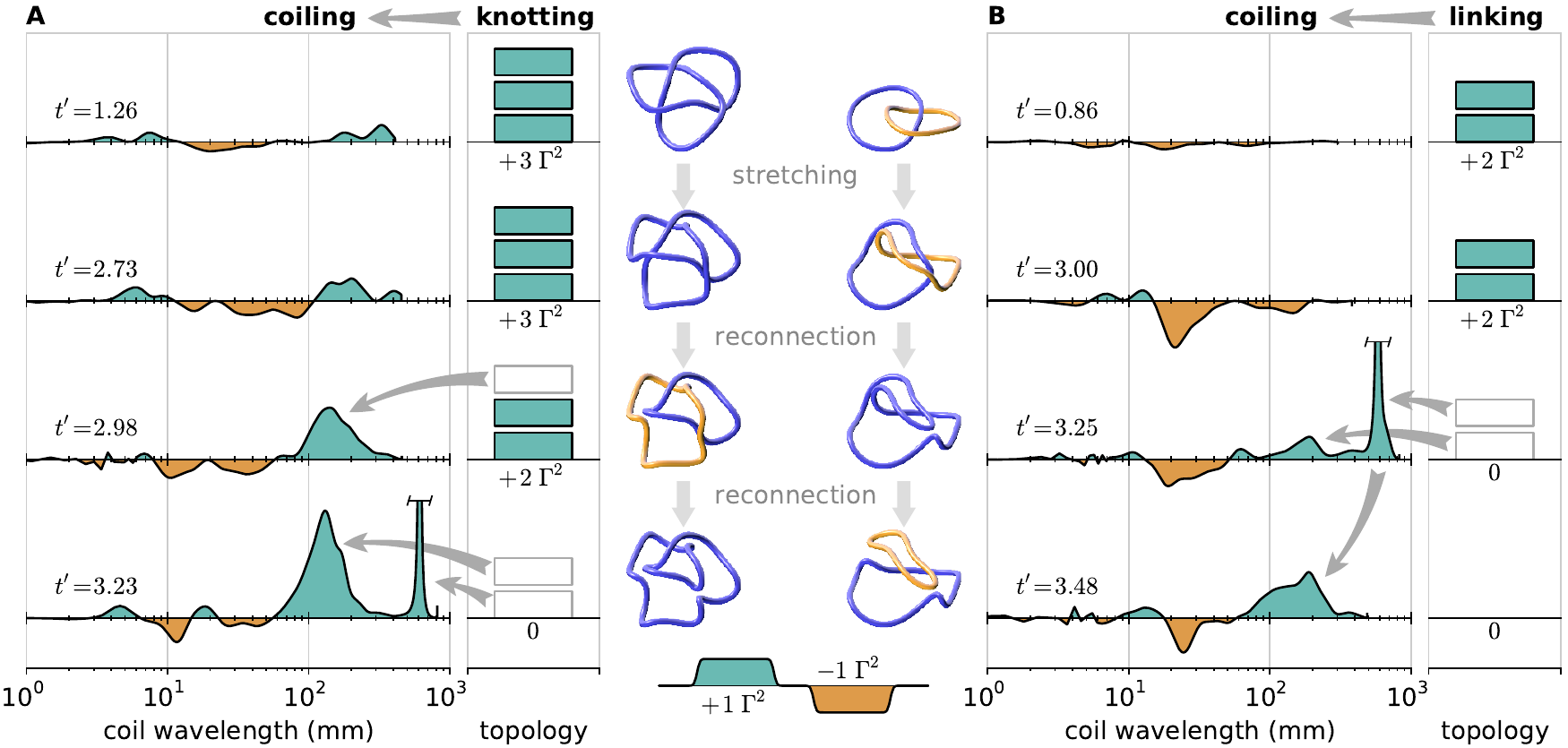}
\caption{
\label{fig:helisto}
(\textbf{A, B}) `Helistograms' for \textbf{A}, a trefoil knot \textbf{B}, linked rings in a viscous fluid experiment (the data set is the same as shown in \figref{fig:Hdata}\textbf{A-D}; the total vortex length is $\sim 1$m for both).  The left portion of each series of plots shows the helicity contribution due to coiling on different spatial scales, obtained by computing $\partial_n \mathcal H_c(\lambda = 10^n)$, where $\lambda$ is the cutoff wavelength for a windowed sinc smoothing.
The right portion of each plot shows the irreducible contribution to the helicity originating from the global vortex topology.  Both the coiling and topological contributions are scaled so that the total helicity is proportional to the filled area of the plots.  The center column shows images of the numerically traced vortices smoothed to $\lambda = 100$ mm.
}
\end{figure*}

The apparent absence of a helicity jump indicates that the vortices are spontaneously arranging themselves into a geometry which allows centerline helicity to be conserved through reconnections.
This proceeds via a simple geometric mechanism: at the moment of topology-changing reconnection, the re-association of vortex lines creates writhing coils in regions that were previously free of writhe, thus converting centerline helicity from linking to writhe (or vice-versa) each time a reconnection takes place.
The remarkable efficiency of the helicity transfer  results from the precise way in which the curves approach each other.
The vortex sections where the reconnection is taking place are almost perfectly anti-parallel just prior to the reconnection event (\figref{fig:reconnection}{\bf C}).
This means that the re-association of vortex tubes that occurs during the reconnection  will not  change the crossing number in \emph{any} projection of the vortex tube centerline. Because the writhe can be computed as the average crossing number over all orientations, this implies the total linking and writhing, $\sum Lk + \sum Wr$, should be conserved, and hence the centerline helicity as well.
(See the supplemental materials for a description of this mechanism purely in terms of planar link-diagrams.)
Alternatively, one can consider the helicity density in the reconnecting region, obtained by computing the tangential flow, $h = \Gamma u_t$ (\figref{fig:Hdata}{\bf E}).
If the annihilated sections are close and anti-parallel, the sum of this helicity density should approach zero, conserving helicity \cite{Moffatt1969a}.
Interestingly, this anti-parallel configuration is expected to form naturally if the vortex tubes are stretching themselves while conserving energy, which seems to happen spontaneously for vortices whose tube topology is non-trivial \cite{Kleckner2013}.

\section{Conversion of Linking and Knotting to Coiling on Different Scales}

The simple geometric mechanism we find for the conservation of centerline helicity through reconnections implies a transfer of helicity across scales which should be quantifiable.
While volumetric Fourier components have been used as measures of helicity content on different scales for flows with distributed vorticity~\cite{Brissaud1973},   for thin core vortices the distance along the vortex provides a natural length scale.
To quantify the storage of helicity as a function of scale along the vortex filament, we  compute the helicity as a function of smoothing, $\mathcal{H}_c(\lambda)$, where $\lambda$ is a hard spatial cutoff scale introduced by convolving the vortex path with a sinc kernel of variable width (this is the same procedure used to smooth the raw data, described above, but with varying cutoff).
When this smoothing is applied, helix-like distortions of the path with period less than $\lambda$ will be removed, and so the contribution of those helical coils to the overall helicity is also removed.
The derivative of this function, $\partial \mathcal H_c |_{\lambda}$, then quantifies the helicity content stored at spatial scale $\lambda$ (see \figref{fig:helistoexample} and supplemental movies S1-2).

Ultimately, there is a component of the helicity that is not removed by even long-scale smoothing; for the relatively simple topologies studied here  the resulting writhe is nearly integer.
This integer component arises because as it is smoothed, the path becomes nearly planar and the integer contribution corresponds to the crossing number  in this effective planar projection.
We refer to this as an effective integer knotting number, akin to linking, and the component removed by smoothing as `coiling', which is produced by helical distortions.

Figure \ref{fig:helisto} shows the helistogram for our trefoil knots and linked rings before, during, and after the reconnection process (see also supplemental movie S3).
In both cases we observe that the initial deformation accompanying the stretching produces small helical deformations across a range of scales; in the case of the trefoil knot these are nearly perfectly balanced, while the linked rings create a strong helix at a scale of 20-30 mm with a helicity opposed to the overall linking.
During the reconnections, we see an immediate transfer from knotting or linking to coiling.
In the case of the linked rings, the first reconnection creates an unlinked geometry immediately (\figref{fig:helisto}{\bf B}, $t'=3.25$), but in doing so creates a large scale folded coil with a spatial scale of $\sim$ 600 mm which then quickly reconnects to form coils at 100--200 mm  (see also \figref{fig:helistoexample}).
The trefoil knot (\figref{fig:helisto}{\bf A}, supplementary movie S4) has similar dynamics; although it is still topologically nontrivial after the first reconnection (\figref{fig:helisto}{\bf A}, $t'=2.98$), it becomes a pair of linked rings which unlinks in a similar manner to the initially linked rings.

In both cases we find that helicity stored on long spatial scales, whether they be knots or links, appears to be intrinsically unstable, cascading through reconnections to smaller spatial scales.
Moreover, this process coincides with the overall stretching that occurs when the topology is non-trivial; after the reconnections take place the length of the vortices appears to stabilize.

Related mechanisms for helicity conservation through reconnections have been suggested in simple models of dissipative plasmas, for example by converting linking to internal twist~\cite{Pfister1991, Lau1996, Berger2006}.

\section{Helicity Conservation in Simulated Superfluids}

\begin{figure}
\includegraphics{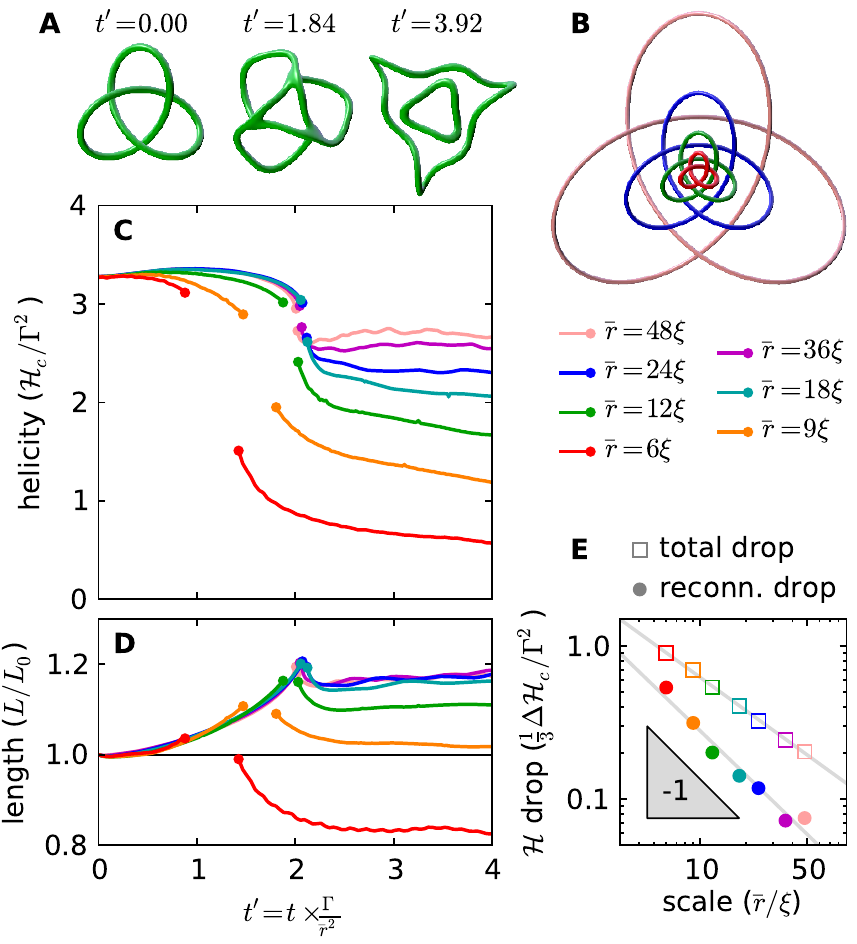}
\caption{
\label{fig:Hgpe}
({\bf A}) Renderings of density iso-surfaces ($\rho = 0.5 \rho_0$) for a trefoil vortex knot ($\bar{r}  = 12 \xi$), simulated with the GPE.  The initially knotted configuration changes to a pair of unlinked rings whose writhe conserves most of the original helicity.
({\bf B}) Renderings of different sized trefoil knots, where the tube radius is given by the healing length, $\xi$, which acts as an effective core size for the superfluid vortex.
({\bf C, D}) The computed centerline helicity ($\mathcal H_c$) and length for of a range of GPE-simulated trefoil knots.  The data is only shown when the distance between vortex lines is $r_{min} > 2 \xi$.
({\bf E})  The helicity jump per reconnection event as a function of size ratio for GPE simulated trefoil knots.  The open squares are the total drop between $t'=0$ and $t'=4$, while the circles indicate the drop during the reconnection event, defined as the missing region in {\bf C} where the vortex tubes overlap.  Larger knots, relative to $\xi$, are found to conserve helicity better by either measure.
}
\end{figure}

The mechanism we observe for helicity conservation through reconnections is entirely geometric, suggesting it may be present in other fluid-like systems as well.
To test this possibility, we simulate the evolution of vortex knots in a superfluid with the Gross-Pitaevskii equation (GPE)~\cite{Pitaevskii2003}.
Although superfluids are inviscid, they are not ideal Euler fluids; thus vortex reconnections are possible and vortex topology is not conserved.
Unlike a classical fluid, which have finite vortex cores, superfluid vortices are confined to a line-like phase defect~\cite{Onsager1949,Feynman1955}; here we track the  centerline helicity (\eqnref{eqn:Hc}) computed for the phase defect path.

Recently, methods have been demonstrated for creating vortex knots in superfluid simulations~\cite{Proment2012}.
We have extended this technique to create initial states with phase defects of arbitrary geometry using velocity integration for flow fields generated by the Biot-Savart law (see supplemental materials for details), allowing us to generate vortices with the same initial shape as our experiments.
We simulate the subsequent vortex evolution using standard split-step methods, and extract the vortex shape by tracing the phase-defects in the resulting wavefunction.
We model vortices with initial radii of $\bar r = 6$--$48 \xi$, where $\xi$ is the healing length which sets the size of a density-depleted region which surrounds the vortex line.  (Our simulations use a uniform grid-size of $0.5 \xi$ and range in resolution from $64^3$ to $512^3$.)

As has been previously observed, we find that vortex knots are intrinsically unstable in superfluids, undergoing a topological and geometrical evolution qualitatively similar to our experimental data (\figref{fig:Hgpe}, see supplemental figure S2 for a corresponding helistogram).
In particular, we find that the reconnections are heralded by an overall stretching of the vortex which abruptly stops after the reconnections take place.
Unlike the experimental data, we also observe a discrete jump in the centerline helicity across the reconnection, ranging from $\Delta \mathcal{H}_c \sim $0.1--1$\Gamma^2$ per reconnection, which is a strong function of the initial vortex size (note that in the simulations all three reconnections happen simultaneously).
If we compute the helicity jump just across the reconnection event, defined as the time during which the colliding vortices are less than 2$\xi$ apart (i.e. when the density depleted regions have already merged), we find a clear $\Delta \mathcal{H}_c \sim \bar r^{-1.0}$ trend.
We observe slightly different results when considering the overall drop in helicity across the entire simulation, from $t'=$0--4, consistent with $\Delta \mathcal{H}_c \sim \bar r^{-0.7}$ trend, which may be a combination of pre-reconnection deformation effects and the reconnection jump.
In particular, the pre-reconnection helicity data seems to converge for $\bar r \gtrsim 18 \xi$, suggesting the finite-core size is not important in this regime prior to the reconnection.

We attribute the loss of centerline helicity to the fact that the
finite size of the depleted density core in the GPE simulations leads
the reconnections to  begin before the vortices are perfectly anti-parallel, resulting in less efficient conservation.
It is unclear if the same effect is present in the experiments, due to the difficulties associated with accurately tracking significantly smaller vortices, however, we note that the expected core-to-vortex size ratio for our experiments is close to that of the largest GPE simulations and we do not observe such a jump.
Previous studies of the simulated dynamics of reconnections in super-fluids and classical fluids suggest that there may be differences in the details of the reconnection behavior~\cite{Kerr2011,Donzis2013,Paoletti2008}.
Nonetheless, we observe that the same conversion of linking and knotting to coiling is present in our superfluid model, indicating that the geometric mechanism for helicity transport across scales that we find in experiment  is generic.

\section{Writhe to Twist Conversion}
\begin{figure}
\includegraphics{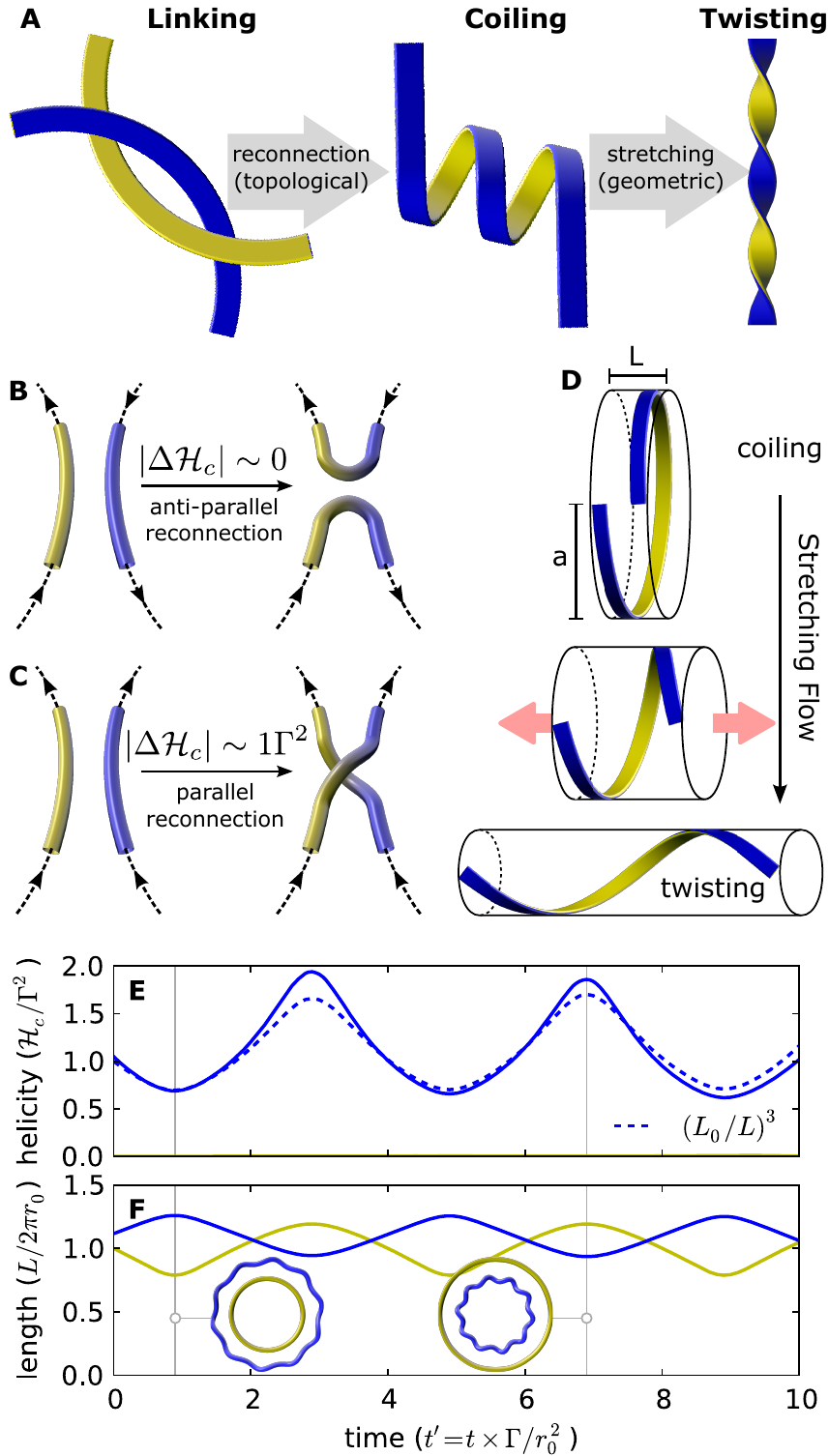}
\caption{
\label{fig:Htransfer}
({\bf A}) Illustrations of mechanisms for storing helicity on different spatial scales; in each case the helicity of the depicted region is the same, $\mathcal H_c = 2 \Gamma^2$.  While linking is global in nature, both coiling and twisting are local -- they produce linking between different sub-sections of the vortex tube, or in this case different edges of the illustrated ribbon.
({\bf B, C}) Diagrams of reconnection events in locally anti-parallel or parallel orientations.  The anti-parallel reconnection does not change helicity because it does not introduce a new `crossing' of the projected tubes, unlike the parallel reconnection.  This anti-parallel configuration tends to form spontaneously for topologically non-trivial vortices, even in the absence of viscosity, in which case helicity is efficiently converted from global linking to local coiling.
({\bf D}) Coiling can be converted to twisting by stretching helical regions of the vortex; this mechanism conserves total helicity because it does not change the topology, but results in an apparent change of helicity when twist can not be resolved.
({\bf E, F}) The helicity (\textbf{E}) and length (\textbf{F}) as a function of time for a simulated geometrical evolution of a circular vortex ring (yellow) `leap-frogging' a vortex ring with a helix superimposed (blue).
}
\end{figure}

Although helicity change is usually understood as being associated with topological changes, we also observe a gradual change in centerline helicity even when reconnections are not taking place, for example in the experimental linked rings (\figref{fig:Hdata}{\bf C}) or the GPE simulations prior to a reconnection (\figref{fig:Hgpe}{\bf C}).
Because the centerline topology is not changing, this helicity change must be attributable to a geometric effect: coiling must have been dissipated or converted to internal twist, which we do not resolve (or is not present, in the case of the GPE).

A dramatic example of this effect can be seen in `leap-frogging' vortices, where a pair of same-sized vortices placed front to back repeatedly passes through one another.
Although this configuration has no centerline helicity if both vortices are perfect rings, a helical winding can be added to one of the rings to give the structure non-zero writhe and hence non-zero centerline helicity.
We model the time evolution of this structure as a thin-core vortex using a simple inviscid Biot-Savart model (see supplemental methods for details), and find that the helicity varies widely as the vortices repeatedly pass through one another (\figref{fig:Htransfer}{\bf EF} and supplementary movie S6).

This variation of the helicity is caused by the fact that the vortices are stretching and compressing each other as a function of time: whenever one vortex passes through another it must be shrunk to fit inside, resulting in a change of the helix pitch.
A simple model of this change can be constructed by first noting that the writhe of a straight helical section is: $Wr_{\rm helix} = N (1 - \cos \theta)$, where $N$ is the number of turns and $\theta$ is the pitch angle.
In the limit of a small pitch angle: $Wr_{\rm helix}\approx N^3 \frac{2 \pi^2 a^2}{L^2}$, where $a$ and $L$ are the radius and length of the cylinder around which the helix is wound (see \figref{fig:Htransfer}{\bf D} and the supplemental materials).
If the flow is a uniform, volume conserving strain, the number of turns is conserved and $a \propto L^{-1/2}$, resulting in $\mathcal{H}_c /\Gamma^2 = Wr \propto L^{-3}$.
This simplistic model qualitatively captures the  centerline helicity of the leap-frogging helix of  \figref{fig:Htransfer}{\bf E}, indicating that the centerline helicity changes primarily because the helical ring is being stretched and compressed.
In general, we expect geometric deformations of the vortex -- including stretching -- should result in continuous changes of the centerline helicity.

Does this imply helicity is not conserved even when the topology is constant?
As discussed in the introduction, helicity can also be stored in twist of the vortex bundle, which is neglected in the centerline helicity.
As a method for keeping track of the vortex bundle orientation, consider it is a ribbon.
If we imagine wrapping this ribbon around a cylinder $N$ times, we expect the linking between the edges of the ribbon to remain constant even if the cylinder changes shape (\figref{fig:Htransfer}{\bf D}).
In this case, the total helicity is constant: $\mathcal{H}/ \Gamma^2 = N  = Wr_{\rm helix} + Tw$
(This is a restatement of the C\u{a}lug\u{a}reanu-White-Fuller Theorem~\cite{Dennis2011,Moffatt1992}).
As the writhe contribution varies dramatically as the vortex is stretched, we conclude that twist must be created in the vortex core to compensate.

We expect similar conversion of writhe to twist is happening for stretching knots and links, although the non-uniform stretching present there produces a rich structure across scales, as seen in our helistograms (\figref{fig:helisto}).
For experimental vortices, the compensating twist should be present but we are not able to directly resolve it; doing so is a challenging goal for future investigations.
As previously noted, however, this twist should be dissipated relatively rapidly by viscosity if the core is small compared to the overall vortex dimensions.
In the case of GPE simulated vortices, the helicity smoothly varies prior to the reconnections -- when rapid stretching is present -- and since there is no method for storing twist it is simply lost.
However, after the reconnections the length and helicity stabilize, despite the fact that the vortices have large oscillating coils.  (The smallest vortices show a slow decay of helicity after reconnection because short wavelength coils, $\lambda \sim \xi$, are radiated away as sound waves in the GPE).

\section{Geometric and Topological Mechanisms for Helicity Conservation}

Our results show that helicity can be conserved in real fluids even when vortex topology is not, and that helicity may not be conserved even when tube topology is invariant.
Vortex reconnections do not simply dissipate helicity, but rather mediate a flow from knotting and linking to coiling, typically from large scales to  smaller scales (\figref{fig:Htransfer}).
The efficient conversion of helicity through a reconnection is due to the anti-parallel vortex configuration (\figref{fig:Htransfer}{\bf B}) that forms naturally in our reconnecting vortices.
Deformation of vortices may also convert coiling into twist on even smaller scales, where it may ultimately be dissipated.
Interestingly, stretching plays a critical role in both topological and non-topological mechanisms for helicity transport, and is also observed to happen spontaneously for initial linked or knotted vortices.
The mechanisms for helicity transport, from linking to coiling to twisting, all have a natural interpretation in terms of the field-line geometry, and as such these mechanisms may play an important role in  any tangled physical field.
Taken as a whole, our results suggests that helicity may yet be a fundamental conserved quantity,
guiding the behavior of dissipative complex flows, from braided plasmas to turbulent fluids.

\begin{acknowledgments}
The authors acknowledge the MRSEC Shared Facilities at the University of Chicago for the use of their instruments.
This work was supported by the National Science Foundation (NSF) Materials Research and Engineering Centers (MRSEC) Program at the University of Chicago (DMR-0820054).
W.T.M.I. further acknowledges support from the A.P. Sloan Foundation through a Sloan fellowship, and the Packard Foundation through a Packard fellowship.
\end{acknowledgments}

\end{document}